\documentclass[footinbib,twocolumn,english,aps,prb,showpacs,superscriptaddress,floats,amsmath,amssymb,floatfix]{revtex4}

\usepackage[latin9]{inputenc}
\usepackage{color}
\usepackage{amsmath}
\usepackage{graphicx}
\usepackage{graphics}
\usepackage{gensymb}
\usepackage{subfigure}
\usepackage{pdftexcmds}
\usepackage{ifpdf}
\usepackage{amssymb}
\usepackage{dsfont}

\makeatletter
\@ifundefined{textcolor}{}
{%
 \definecolor{BLACK}{gray}{0}
 \definecolor{WHITE}{gray}{1}
 \definecolor{RED}{rgb}{1,0,0}
 \definecolor{GREEN}{rgb}{0,1,0}
 \definecolor{GREEN2}{rgb}{0,0.4,0}
 \definecolor{BLUE}{rgb}{0,0,1}
 \definecolor{CYAN}{cmyk}{1,0,0,0}
 \definecolor{MAGENTA}{cmyk}{0,1,0,0}
 \definecolor{YELLOW}{cmyk}{0,0,1,0}
 \definecolor{YELLOW2}{cmyk}{0,0,1,0.6}
 \definecolor{ORANGE}{rgb}{1,0.22,0}

 }

\def \be {\begin{equation}}
\def \ee {\end{equation}}
\def \bea {\begin{eqnarray}}
\def \eea {\end{eqnarray}}
\DeclareMathOperator\Tr{Tr}
\allowdisplaybreaks
\makeatother

\begin{document}

\title{From skyrmions to $\mathds{Z}_{2}$ vortices in distorted chiral antiferromagnets}

 \author{S.A. Osorio}
 \affiliation{Instituto de F\'isica de L\'iquidos y Sistemas Biol\'ogicos, CCT La Plata, CONICET and Departamento de F\'isica, Facultad de Ciencias Exactas, Universidad Nacional de La Plata, C.C. 67, 1900 La Plata, Argentina}
 \author{M.B. Sturla}
\affiliation{Instituto de F\'isica de L\'iquidos y Sistemas Biol\'ogicos, CCT La Plata, CONICET and Departamento de F\'isica, Facultad de Ciencias Exactas, Universidad Nacional de La Plata, C.C. 67, 1900 La Plata, Argentina}
 \author{H.D. Rosales}
 \affiliation{Instituto de F\'isica de L\'iquidos y Sistemas Biol\'ogicos, CCT La Plata, CONICET and Departamento de F\'isica, Facultad de Ciencias Exactas, Universidad Nacional de La Plata, C.C. 67, 1900 La Plata, Argentina}
\affiliation{Departamento de Cs. B\'asicas, Facultad de Ingenier\'ia, Universidad Nacional de La Plata, C.C. 67, 1900 La Plata, Argentina}
 \author{D.C. Cabra}
 \affiliation{Instituto de F\'isica de L\'iquidos y Sistemas Biol\'ogicos, CCT La Plata, CONICET and Departamento de F\'isica, Facultad de Ciencias Exactas, Universidad Nacional de La Plata, C.C. 67, 1900 La Plata, Argentina}
 \affiliation{Abdus Salam International Centre for Theoretical Physics, Associate Scheme, Strada Costiera 11, 34151, Trieste, Italy}
\begin{abstract}
Swirling topological spin configurations, known as magnetic skyrmions, are known to be stabilised in chiral ferromagnets with Dzyaloshinskii-Moriya interaction (DMI). In particular, for appropriate values of the external magnetic field they appear in a topological crystalline phase, termed Skyrmion crystal phase (SkX). A similar phenomenon is present in the antiferromagnetic case, for the Heisenberg triangular antiferromagnet (HTAF) with DMI. Here, the most striking feature is that  the emergent topological phase consists of three SkX interpenetrated  sublattices. On the other hand, the pure HTAF, being described by an SO(3) order parameter, can host $\mathds{Z}_{2}$ vortices. This rises the fundamental question on whether both non trivial structures are related.  In this paper we unravel a hidden connection between both topological entities by studying the HTAF with {\it anisotropic} DMI. To this end, we combine an effective field theory description, the Luttinger-Tisza approximation and Monte Carlo simulations. 
We show that even a slight anisotropy in the DMI proves to be the key ingredient to deform the interpenetrated SkX structure and reveal a $\mathds{Z}_2$-vortex crystal.
\end{abstract}

\maketitle

{\it Introduction.--} Skyrmions are topological magnetic units which, either in isolated form or organized in a crystal structure (SkX) \cite{BY_89,BH2_94,MBJ_09,YOK_10,Han_10}, are by now ubiquitous in nature.  In most  cases, they are stabilized in ferromagnetic systems with a combination of an external magnetic field and  DMI\cite{NagaosaTokura2013,finocchio_2016,fert_2017,everschor_2018}. In the case of the HTAF, the periodic magnetic structure shows up in a more intricate way: it arises as three different SkX's, one on each sublattice, which interpenetrate and lead to a new crystal state (see Refs. [\onlinecite{Rosales_15,OSO_17}]) that was termed AF-SkX (see also Ref. [\onlinecite{diaz2019topological}]). The nature of this intricate magnetic structure is the main object of study in this letter. 

In the case of zero DMI, the HTAF at finite temperature has been studied in Ref. [\onlinecite{KAWA_MIYA_1984}] where it was shown that, in connection to the SO(3) structure of the order parameter, a binding-unbinding transition of $\mathds{Z}_{2}$ vortices takes place. More recently, Rousochatzakis et al [\onlinecite{daghofer_16}] studied the HTAF with Kitaev-like interactions and found an intermediate phase in which $\mathds{Z}_{2}$ vortices crystallize in a triangular array (we call it here $\mathds{Z}_{2}$X state).
Interestingly, we have found that also the AF-SkX structure can lead to a periodic arrangement of pairs of $\mathds{Z}_{2}$ vortices:
we show that the key ingredient that makes connection between these two topologically different phases is an anisotropic DMI. 

More precisely, in this Letter we show that spatial anisotropy in the DMI
within the HTAF (whose origin could be structural or artificially induced) can be used as a tuning parameter that allows one to move from a  topologically smooth non-trivial configuration, the AF-SkX, to a topologically singular one, a lattice of pairs of $\mathds{Z}_{2}$ vortices.

In order to show this, we first explore the possible modulated states through the Luttinger-Tisza approximation (LTA)\cite{luttinger1946theory} and then study the magnetic phase diagram using Monte Carlo (MC) simulations. Then, we derive an effective field theory (ET) which provides further support to the numerical results. We focus our attention on the emergence of a complex magnetic phase described as a $\mathds{Z}_{2}$ vortex-pair lattice, that in each sublattice presents a periodic arrangement of deformed skyrmions. 

\begin{figure*}
\includegraphics[width=15cm]{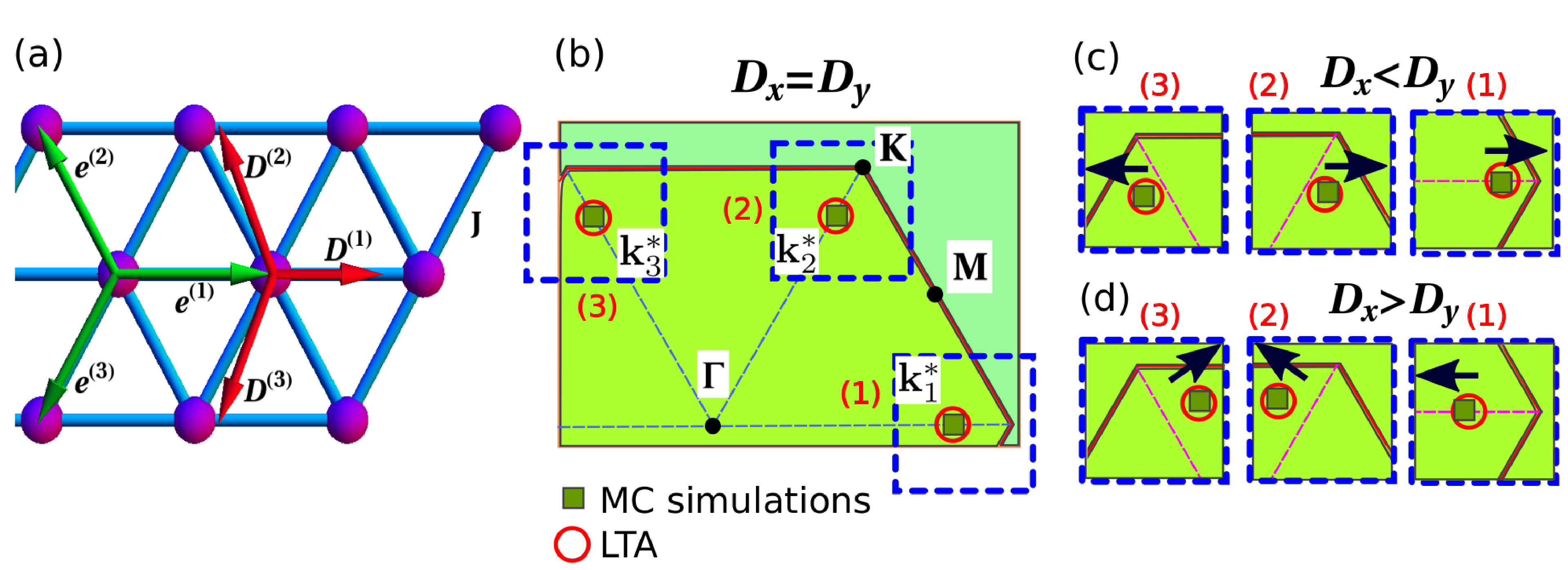}
\caption{(Color online) (a): Triangular lattice. The red arrows indicate the direction of the anisotropic DMI vectors, which depend on two parameter which control the spacial anisotropy: ${\bf D}^{(1)}=(D_{x},0)$, ${\bf D}^{(2)}=(-D_{x}/2,\sqrt{3}D_{y}/2)$ and ${\bf D}^{(3)}=(-D_{x}/2,-\sqrt{3}D_{y}/2)$. The green solid arrows are unit vectors.  Ordering vectors ${\bf k}^*_i$ for   (b) $D_x=D_{y}=0.35\,J$, (c) $D_{x}=0.27\,J$ and $D_{y}=0.35\,J$ and (d) $D_{x}=0.35\,J$ and $D_{y}=0.27\,J$. The green squares correspond to the Bragg peaks of the in plane structure factor ($\mathbf{S}_{\bot}(\mathbf{k})$) calculated from 100 MC snapshots at $B/J=2.7$ and $T/J\approx1\times10^{-2}$. The red circles indicates the set of minima of $\lambda^{-}(\mathbf{k})$. The red hexagon delimit the first Brillouin zone of the triangular lattice. The black arrows represent qualitatively the displacement of the minima of $\lambda^{-}(\mathbf{k})$ and the peaks of $\mathbf{S}_{\bot}(\mathbf{k})$ with respect to the isotropic case.}
\label{fig:SPH_FAC} 
\end{figure*}

{\it Model.--} We start by considering the following microscopic Hamiltonian for a classical spin system on a triangular two-dimensional lattice:
\be
H=\sum_{<\mathbf{r},\mathbf{r'}>}\left[J(\mathbf{S}_{\mathbf{r}}\cdot\mathbf{S}_{\mathbf{r'}})+\mathbf{D}_{\mathbf{r},\mathbf{r'}}\cdot(\mathbf{S}_{\mathbf{r}}\times\mathbf{S}_{\mathbf{r'}})\right]-\mathbf{B}\cdot\sum_{\mathbf{r}}\mathbf{S}_{\mathbf{r}},
\label{eq:Hamil}
\ee
where the spin is a unimodular vector, $\left|\mathbf{S}_{\mathbf{r}}\right|=1$, $J>0$ is the antiferromagnetic exchange coupling and $\mathbf{D}_{\mathbf{r},\mathbf{r'}}$ are vectors describing the DMI. In this work we will consider the effect of spacial anisotropy in the DMI, controlled through two parameters $D_x,D_y$ as shown in Fig. \ref{fig:SPH_FAC}(a).

We start by looking for possible modulated states at zero temperature and zero magnetic field by means of the LTA. The spectrum obtained within this method consists of three bands $\lambda^{-}(\mathbf{k})$, $\lambda^{0}(\mathbf{k})$ and $\lambda^{+}(\mathbf{k})$. For $D_{x}=D_{y}=0$,  the three overlap completely ($\lambda^{+}(\mathbf{k})=\lambda^{0}(\mathbf{k})=\lambda^{-}(\mathbf{k})$) and the ordering corresponds to a $120\degree$ structure with the wave vector $\mathbf{k}=(4\pi/3,0)$. 
For nonzero DMI (either $D_{x,y}\neq 0$ or both) there is a minimum surface $\lambda^{-}(\mathbf{k})$ which does not intersect the other bands and hence the magnetic order is determined by its minima ${\bf k}^{*}$. There exist three such non-equivalent local minima ${\bf k}^{*}_i$ indicated in Fig.\ref{fig:SPH_FAC}. 
For the isotropic case the minima lie on the vertices of a regular hexagon and the three minima are equivalent\cite{Rosales_15,OSO_17} having the same energy. In the anisotropic case $D_x\neq D_y$ the minima are distributed along an irregular hexagon. In this case we have $\lambda^{-}({\bf k}^{*}_{1})\neq\lambda^{-}({\bf k}^{*}_{2})=\lambda^{-}({\bf k}^{*}_{3})$.
If $D_{x}<D_{y}$ then $\lambda^{-}({\bf k}^{*}_{1})>\lambda^{-}({\bf k}^{*}_{2,3})$, if $D_{x}>D_{y}$ then $\lambda^{-}({\bf k}^{*}_{1})<\lambda^{-}({\bf k}^{*}_{2,3})$. 
This result expresses the possible existence of a phase with a modulated magnetic order with a definite ${\bf k}^{*}$, the one for which $\lambda^{-}$ takes the absolute minimum. However, since the difference between the energy of the absolute minimum and the local one is small ($\approx 1\%$), the presence of the external magnetic field and temperature could mix them all, stabilizing a triple-q state.

{\it Monte Carlo simulations.--}The presence of this triple-q phase (at finite magnetic field and temperature) without the restrictions imposed by the  LTA approach, can be tested by classical Monte Carlo (MC) simulations of the spin Hamiltonian
given in Eq. (\ref{eq:Hamil}). To this end, we study the parallel and perpendicular (to $z$) components of the structure factor $\mathbf{S}_{\parallel}(\mathbf{k})=\frac{1}{N}\left\langle \left|\sum_{\mathbf{r}}S^{z}_{\mathbf{r}}e^{-i\mathbf{k}\cdot\mathbf{r}}\right|^{2}\right\rangle$, $\mathbf{S}_{\bot}(\mathbf{k})=\frac{1}{N}\left\langle \left|\sum_{\mathbf{r},\alpha=x,y}S^{\alpha}_{\mathbf{r}}e^{-i\mathbf{k}\cdot\mathbf{r}}\right|^{2}\right\rangle$. In Fig. \ref{fig:SPH_FAC} we present the results for $\mathbf{S}_{\bot}(\mathbf{k})$ (the  component $\mathbf{S}_{\parallel}(\mathbf{k})$ has the same behaviour and is not included here) for $B/J=2.7$ and $T/J=10^{-2}$.
We observe the presence of a three-peaks pattern where the minima positions are modified by the anisotropic DMI. The position of these peaks can be compared  with the minima found within the LTA, showing an excellent agreement in the displacement produced by the anisotropy (see Fig. \ref{fig:SPH_FAC}). The similarity between the isotropic and the anisotropic structure factors, suggest that an interpenetrated spin structure could also be present in the anisotropic case\cite{Rosales_15,OSO_17}.  

\begin{figure}[htb]
\includegraphics[width=8cm]{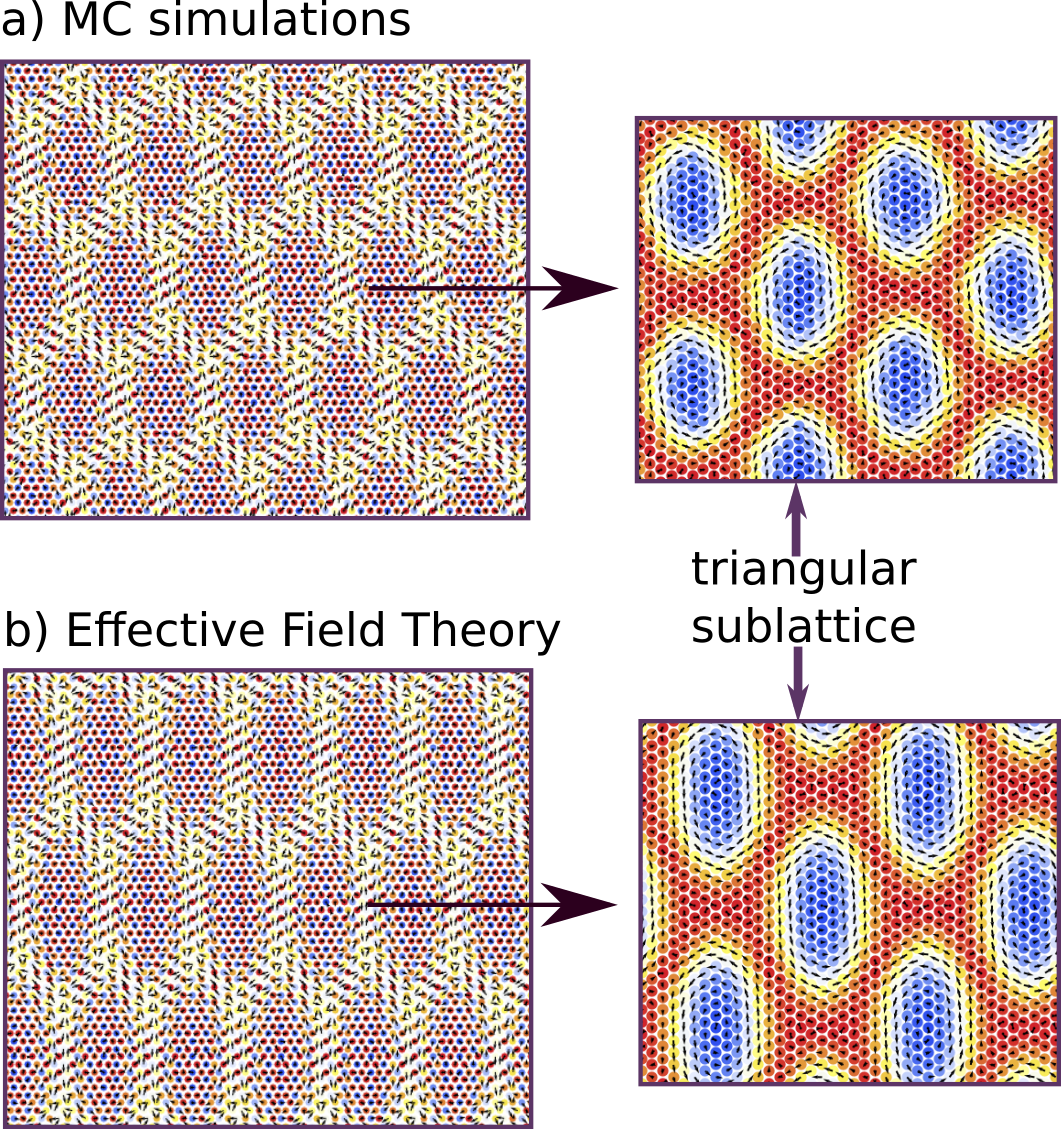}
\caption{(Color online) Spin configuration for the full lattice (left) and for one sublattice (right) for $D_{x}/J=0.35$, $D_{y}/J=0.27$ and $B/J=2.7$ obtained from (a) MC simulations at $T/J\approx10^{-2}$ and (b) the discretization of the spin field from the ET results. The arrows represent the in-plane components of the spin, the colour represents the $S_{z}$ component (blue stands for $S_{z}=-1$ and red for $S_{z}=1$).
}
\label{fig:skyrmions}
\end{figure}

The analysis of the low temperature phase diagram for the anisotropic case provides similar results as the isotropic one. In particular, for intermediate fields $2.2<B/J<5$, the AF-SkX phase is stabilized. The main difference is that, on each interpenetrated lattice, the anisotropy leads to oblate skyrmions (see Fig. \ref{fig:skyrmions}(a)). The skyrmions stretch along the $\mathbf{\hat{y}}$ ($\mathbf{\hat{x}}$) direction for $D_{x}>D_{y}$ ($D_{x}<D_{y}$) and they are located on the vertices of isosceles triangles (instead of equilateral ones, as occurs in the isotropic case).\\

As we show below, the oblate skyrmions arranged in the distorted triangular pattern on each sublattice, produce the emergence of a lattice of $\mathds{Z}_{2}$ vortex pairs.
It is well known that, in the absence of DMI and for very small magnetic fields, the HTAF can be described by an SO(3) order parameter \cite{KAWA_MIYA_1984,DR_89}.
From the SO(3) description, the spin configurations of the triangular antiferromagnetic system can be classified by the first homotopy group $\Pi_{1}(SO(3))\approx\mathds{Z}_{2}$. This means that, from a topological perpective, there only exists two non-equivalent configurations, the trivial (non-singular) configuration and the non trivial (singular) one. 

In order to study the possible appearance of these $\mathds{Z}_{2}$ vortices in the distorted model, we introduce the chirality vector field  $\mathbf{\kappa}$ following \cite{KAWA_MIYA_1984}: 

\be
\mathbf{\kappa}=\frac{2}{3\sqrt{3}}\left(\mathbf{S}_{1}\times \mathbf{S}_{2}+\mathbf{S}_{2}\times \mathbf{S}_{3}+\mathbf{S}_{3}\times \mathbf{S}_{1} \right).
\label{eq:chirality}
\ee
This vector, in the absence of DMI, is perpendicular to the plane of the $120\degree$ structure. Its modulus, $0\leq\left|\mathbf{\kappa}\right|\leq 1$, provides information about the degree of deformation of the $120\degree$ structure, vanishing whenever two spins are parallel and giving $\left|\mathbf{\kappa}\right|=1$ when the $120\degree$  planar configuration is achieved.
In order to calculate the vorticity, we assign a trihedron (rigid-body) to each triangular plaquette associated with their spin configuration. This trihedron is represented by three mutually orthogonal unit vectors constructed from the spin field. We chose these vectors as indicated in Ref. \cite{KAWA_MIYA_1984}: the first vector is chosen to be in the direction of the chirality vector $\mathbf{n}_{1}=\hat{\kappa}$, the second vector is $\mathbf{n}_{2}=(\mathbf{n}_{1}\times\mathbf{S}_{i})\times\mathbf{n}_{1}$ where $\mathbf{S}_{i}$ is the spin vector on a given sublattice, and the third is the cross product of the previous, $\mathbf{n}_{3}=\mathbf{n}_{1}\times\mathbf{n}_{2}$. The rigid-body transforms from a point $p$ to another point $p'$ through a matrix $\mathbf{R}\in$ SO(3).
By following the prescription introduced by Kawamura and Miyashita \cite{KAWA_MIYA_1984} we can assign a unique SU(2) matrix  (this representation is two-to-one, two elements of SU(2) correspond to the same element of SO(3)) to each SO(3)-matrix along the line connecting two adjacent points of the plaquette lattice. We call $\mathbf{U}_{i}$ to the matrix associated to the i-th link. 

The vorticity $\mathcal{V}_{C}$ for a closed path $C$ is defined as \cite{KAWA_MIYA_1984}
\be
\mathcal{V}_{C}=\frac{1}{2}\Tr\left\lbrace \prod_{i\in C} \mathbf{U}_{i} \right\rbrace,
\ee
assuming that the $\mathds{Z}_{2}$ group is represented by the elements $\left\lbrace -1,1 \right\rbrace$ with the usual product, the vorticity takes the value $\mathcal{V}_{C}=-1$ ($\mathcal{V}_{C}=+1$) for trajectories enclosing an odd (even) number of vortices.
The smallest loops on the plaquette triangular lattice correspond to the non-equivalent triangles which we call $C_{\vartriangleleft}$ and $C_{\vartriangleright}$. They connect three adjacent plaquettes and they do not overlap with each other (except for the vertices and edges they could have in common).

\begin{figure}[htb]
\includegraphics[width=8.5cm]{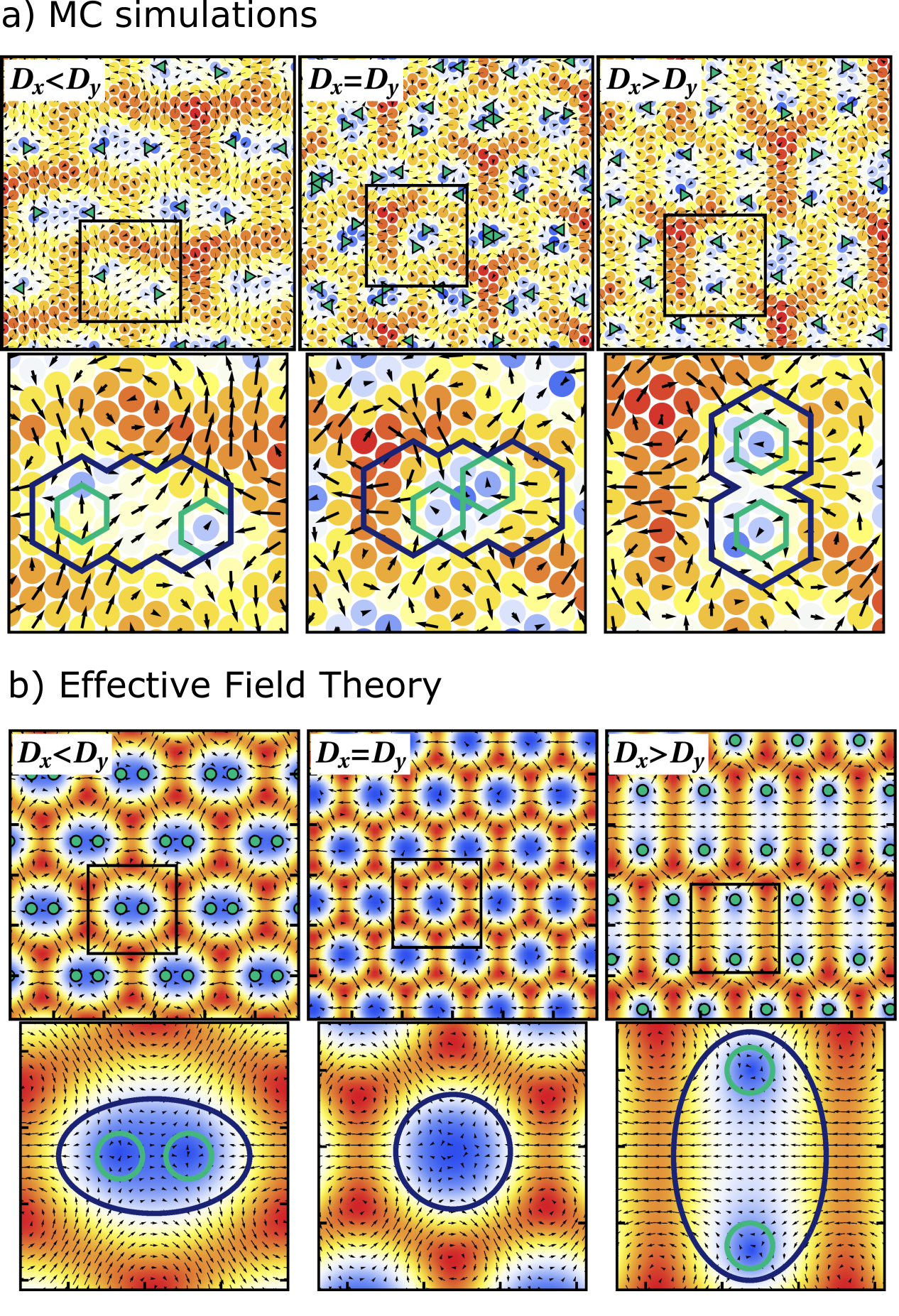}
\caption{(Color online) Vortex configuration for different values of $D_{x}$, $D_{y}$ at $B/J=2.7$. On the left side, $D_y/D_x=1.3$ ($D_y/J=0.35$), the central panel $D_{x}/D_{y}=1$ ($D_x/J=0.35$), and the right panel is for $D_{y}/D_x=0.77$ ($D_{x}/J=0.35$). The second lines correspond to a zoom of the region bounded by the inner black squares. In the zoomed regions the vorticity is $\mathcal{V}=-1$ along the green curves and $\mathcal{V}=1$ for the deep blue curves. (a) Results from MC simulations at $T/J\approx10^{-2}$. The green triangles represent the position of the vortices. (b) Results from ET. The green points indicate the position of the vortices. In the complete graph the distance between the ticks is $8a$ ($a$ is the lattice parameter) and in the zoomed region it is $4a$. The arrows correspond to the in-plane component of chirality vector and the color represents its length, blue (red) for $\left|\mathbf{\kappa}\right|=0$ ($\left|\mathbf{\kappa}\right|=1$).}
\label{fig:VORT}
\end{figure}

In Fig. \ref{fig:VORT} we show the chirality $\kappa$ given by Eq. (\ref{eq:chirality}) on each plaquette of the latttice. As in previous studies\cite{KAWA_MIYA_1984,TREBST_15}, here the $\mathds{Z}_{2}$-vortices are sited over the zeros of the in-plane chirality vector field, with non-trivial index \cite{Milnor1997}.  We present the results for three representative cases: $D_{x}>D_{y}$, $D_{x}=D_{y}$ (isotropic case) and $D_{x}<D_{y}$, from both the MC simulations (Fig. \ref{fig:VORT}(a)) at low temperatures ($T\approx10^{-2}$J) and by means of the analysis of the ET at zero temperature (Fig. \ref{fig:VORT}(b)). The localization of the vortices is determined from MC snapshots by the calculation of the vorticity around the smallest closed curves. We calculate the vorticity for all the $C_{\vartriangleleft}$ and $C_{\vartriangleright}$ triangles. The green triangles are those for which $\mathcal{V}=-1$ indicating the presence of a vortex in that position of the lattice. We recognize pairs of vortices in the regions where the modulus of $\mathbf{\kappa}$ is minimum and the in-plane chirality is zero.

For the isotropic case we can see that the vortices are almost superimposed and the direction of separation is randomly oriented. On the other side, the presence of the anisotropy $D_x\neq D_y$ increases the separation between the vortices. The direction of the separation is determined by the anisotropy: if $D_{y}>D_{x}$ ($D_{x}>D_{y}$) they depart from each other in the $\mathbf{\hat{x}}$ ($\mathbf{\hat{y}}$) direction as shown in Fig. \ref{fig:VORT}(a). The green contours (in the lower panels of Fig. \ref{fig:VORT}(a)) are those for which we count a negative vorticity indicating that this contour surrounds a vortex. For the deep blue loop (which encloses the regions delimited by the two green loops) we have $\mathcal{V}=1$, meaning that it contains an even number of vortices, as expected. 

{\it Effective field theory.--} To further analyze the appearance of the $\mathds{Z}_{2}$X lattice as we turn on the DMI anisotropy, we extend the effective field theory approach (following the same lines as in a previous work\cite{OSO_17}) in order to analyse the effect of the anisotropy in this vortex separation. We consider a set of variational Ans\"atze to describe the AF-SkX phase, with the modification that the $\mathbf{k}_{i}$ ($i=1,2,3$, that span the SkX phase represented as a triple-q phase) are not of the same length and they do not form an angle of 120$\degree$. However they still satisfy $\sum_{i}\mathbf{k}_{i}=\mathbf{0}$, with two vectors of the same length and different from the third: $\left|\mathbf{k}_{1}\right|\neq\left|\mathbf{k}_{2}\right|=\left|\mathbf{k}_{3}\right|$. The terms containing derivatives of $\mathbf{M}$ ($\nabla^{2}\mathbf{M}$, $\partial_{x}\mathbf{M}$ and $\partial_{y}\mathbf{M}$) are negligible for the AF-SkX phase considered here\cite{OSO_17}. Thus the minimal Hamiltonian to describe this phase is given by:

\bea
\nonumber
\mathcal{H}&=&\mathcal{H}_{M}+\sum_{i=1}^{3}\mathcal{H}_{i},\\
\nonumber
\mathcal{H}_{M}&=&\frac{J}{2}(\mathbf{M}^{2}-3),\\
\nonumber
\mathcal{H}_{i}&=&-a^2 \frac{J}{8}\mathbf{S}_{i}\cdot\nabla^{2}\mathbf{S}_{i}-\frac{1}{3}\mathbf{B}\cdot\mathbf{M}-\\
&-&\frac{a}{4}\mathbf{S}_{i}\cdot\left[D_{x}\left(\partial_{x}\mathbf{S}_{i}\times\mathbf{\hat{x}}\right)+
D_{y}\left(\partial_{y}\mathbf{S}_{i}\times\mathbf{\hat{y}}\right)\right] \ .
\label{eq:modminimo}
\eea

With this Hamiltonian and generalizing the Ansatz used in Ref. \cite{OSO_17} for the anisotropic case, we find that for magnetic fields $2.4<B/J<4.5$ a modified SkX phase on each spin sublattice shows up, see Fig. \ref{fig:skyrmions}(b). As in the MC results, these modified SkX sublattice structures can be described as oblate skyrmions arranged on a triangular lattice with one of its sides elongated\footnote{The same configuration of elliptical skyrmions was encountered in the context of chiral ferromagnets with anisotropic DMI \cite{SHI_15, OSO_19}.}.

The green filled circles in Fig. \ref{fig:VORT}(b) represent the positions of the vortices. In the isotropic case the minima of $\left|\mathbf{\kappa}\right|$ do not correspond to vortices, since for every closed curve, we find that the vorticity is $\mathcal{V}=1$, that is, there are no vortices.  However, we can see that a pair of vortices appears when the anisotropy is turned on, as depicted in the left and right panels of Fig. \ref{fig:VORT}(b). As shown in the lower panels, for a loop enclosing only one minimum (green circle) the vorticity is $\mathcal{V}=-1$, while for a loop enclosing two minima (deep blue) the vorticity is $\mathcal{V}=1$. Again the separation of the vortices is determined by the anisotropy taking place along the $\mathbf{\hat{y}}$ ($\mathbf{\hat{x}}$) direction for $D_{x}>D_{y}$ ($D_{y}>D_{x}$). These results coincide with those obtained from MC simulations. From all this, we conclude that pairs of vortices appear in the ground state, forming a triangular lattice, as soon as the DMI anisotropy is turned on. In the isotropic limit these pairs of vortices are superimposed resulting in a topologically smooth configuration, while in the case of anisotropic DMI, the vortices split away, and the configuration becomes topologically singular.  
We should note that within the effective field theory the lattice distance is effectively set to zero while in the MC simulations it is the effect of the lattice distance which produces a mismatch between the interpenetrated skyrmion lattices and hence one observes closely bounded pairs of vortices in the isotropic case.

{\it Outlook.--} In summary, in this manuscript we have studied the connection between the intrincate topological structure present in the AF-SkX phase  and the $\mathds{Z}_{2}$ vortices formed by chirality vectors. To carry out this challenging task, we have resorted to a combination of techniques, comprising an effective field theory description in the continuum limit, the Luttinger-Tisza approximation and extensive Monte-Carlo simulations. 

Our most striking result is that in the AF-SkX phase, composed of three interpenetrated  ferromagnetic skyrmion lattices, a small anisotropy in the DM interaction induces the emergence of a periodic lattice of stable topological $\mathds{Z}_{2}$ vortex-pairs that covers the entire magnetic system. In order to detect this vortex structure, formed by chirality vectors, we compute the vorticity associated with closed contours on the dual lattice. 
We found that the anisotropy  is the key to control the separation between vortices, while keeping the topological structure of the AF-SkX phase stable. We have confirmed this scenario by means of large scale Monte Carlo simulations. A most relevant feature in this new topological phase is that the vortex-pair lattice emerges in the ground state, whereas in the pure HTAF the  $\mathds{Z}_{2}$ vortices  appear in thermally excited states\cite{KAWA_MIYA_1984}. 

This study thus proposes a tool to explore vortex lattices: the manipulation of the DM anisotopy. This opens the door to the exploration of new topological proporties.
We expect that the anisotropy of the interactions may be experimentally realised in thermal strain studies, as done for example in  the FeGe compound \cite{SHI_15}. A challenging goal is the detection of the proposed $\mathds{Z}_2$-vortex lattice, yet it has been suggested that the dynamical spin structure factor, which can be measured by  means of inelastic neutron scattering, could be the key to achieve this objective \cite{okubo2010signature}.

{\it Acknowledgments.--} This work was partially supported by CONICET (PIP 2015-813) and ANPCyT (PICT 2012-1724). H.D.R. acknowledges support from PICT 2016-4083. S.A.O acknowledges to Mart\'in Villalba and Flavia Gomez Albarrac\'in for fruitful discussions.

\bibliography{references.bib}

\end{document}